%% file: maximum_labelled_clique.tex
\documentclass[smallextended]{svjour3}

\usepackage{complexity}                
\usepackage{varioref}                  
\usepackage{hyperref}                  
\usepackage{cleveref}                  
\usepackage{tikz}                      
\usepackage[ruled,vlined]{algorithm2e} 
\usepackage{amsmath}                   
\usepackage{booktabs}                  
\usepackage{siunitx}
\usepackage{wrapfig}
\usepackage{pdflscape}
\usepackage{afterpage}

\usetikzlibrary{arrows, shadows, calc, positioning, decorations, decorations.pathmorphing,
    decorations.pathreplacing, patterns}

\definecolor{safelightblue}{rgb}{0.65098, 0.807843, 0.890196}
\definecolor{safedarkblue}{rgb}{0.121569, 0.470588, 0.705882}
\definecolor{safeverylightblue}{rgb}{0.857843, 0.931961, 0.996078}
\definecolor{safeverydarkblue}{rgb}{0.007843, 0.219608, 0.345098}
\definecolor{safelightorange}{rgb}{0.996078, 0.931961, 0.857843}
\definecolor{safelightgreen}{rgb}{0.698039, 0.87451, 0.541176}
\definecolor{safemediumorange}{rgb}{0.74902, 0.505882, 0.490196}
\definecolor{safenearlywhite}{rgb}{0.9, 0.9, 0.9}

\crefname{algocf}{Algorithm}{Algorithms}
\Crefname{algocf}{Algorithm}{Algorithms}
\crefname{figure}{Figure}{Figures}
\Crefname{figure}{Figure}{Figures}
\crefname{table}{Table}{Tables}
\Crefname{table}{Table}{Tables}

\makeatletter
\def\cl@chapter{\@elt {theorem}}
\makeatother

\newcommand{\Cbest}{C^\star}
\newcommand{\Lbest}{L^\star}
\newcommand{\first}{\mathit{first}}
\newcommand{\bounds}{\mathit{bounds}}
\newcommand{\budget}{\mathit{budget}}
\newcommand{\order}{\mathit{order}}
\newcommand{\colour}{\mathit{colour}}
\newcommand{\uncoloured}{\mathit{uncoloured}}
\newcommand{\colourable}{\mathit{colourable}}



\journalname{Optimization Letters}

\begin{document}

\title{A Parallel Branch and Bound Algorithm for the Maximum Labelled Clique Problem}
\titlerunning{The Maximum Labelled Clique Problem}

\author{
    Ciaran McCreesh%
    \thanks{This work was supported by the Engineering and Physical Sciences Research Council [grant number EP/K503058/1]}
    \and
    Patrick Prosser
}

\institute{C. McCreesh and P. Prosser \at University of Glasgow \\ Glasgow, Scotland \\
    \email{c.mccreesh.1@research.gla.ac.uk} and \email{patrick.prosser@glasgow.ac.uk}}

\date{Received: date / Accepted: date}

\maketitle

\begin{abstract}
    The maximum labelled clique problem is a variant of the maximum clique problem where edges in
    the graph are given labels, and we are not allowed to use more than a certain number of distinct
    labels in a solution. We introduce a new branch-and-bound algorithm for the problem, and explain
    how it may be parallelised.  We evaluate an implementation on a set of benchmark instances, and
    show that it is consistently faster than previously published results, sometimes by four or five
    orders of magnitude.

\end{abstract}

\section{Introduction}

\begin{wrapfigure}[14]{R}{0.45\textwidth}
    \centering
    \begin{tikzpicture}
        \input{figure-clique}
    \end{tikzpicture}
    \caption{A graph with maximum clique $\{1, 2, 3, 4, 5\}$, using all four edge labels. If our
    budget is only three, a maximum feasible clique has size four. There are several such cliques,
    but $\{4, 5, 6, 7\}$ is optimal since it uses only two labels, whilst every other
    uses at least three.}
    \label{fig:clique}
\end{wrapfigure}
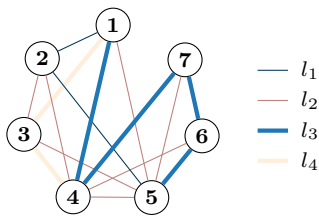

A clique in a graph is a set of vertices, where every vertex in this set is adjacent to every other
in the set. Finding the size of a maximum clique in a given graph is one of the fundamental \NP-hard
problems. Carrabs et al.\ \cite{Carrabs:2014} introduced a variant called the maximum labelled
clique problem. In this variant, each edge in the graph has a label, and we are given a budget $b$: we
seek to find as large a clique as possible, but the edges in our selected clique may not use more
than $b$ different labels in total. In the case that there is more than one such maximum, we must
find the one using fewest different labels. We illustrate these concepts in \cref{fig:clique}, using
an example graph due to Carrabs et al.; our four labels are shown using different styled edges.

Carrabs et al.\ give example applications involving social network analysis and telecommunications.
For social network analysis, vertices in the graph may represent people, and labelled edges describe
some kind of relationship such as a shared interest. We are then seeking a large, mutually connected
group of people, but using only a small number of common interests. For telecommunications, we may
wish to locate mirroring servers in different data centres, all of which must be connected for
redundancy. Labels here tell us which companies operate the connections between data centres: for
simplicity and cost, we have a budget on how many different companies' connections we may use.

A mathematical programming approach to solving the problem was presented by Carrabs et al., who used
CPLEX to provide experimental results on a range of graph instances. Here we introduce the first
dedicated algorithm for the maximum labelled clique problem, and then describe how it may be
parallelised to make better use of today's multi-core processors. We evaluate our implementation
experimentally, and show that it is consistently faster than that of Carrabs et al., sometimes by
four or five orders of magnitude. These results suggest that state of the art maximum clique
algorithms are not entirely inflexible, and can sometimes be adapted to handle side constraints and
a more complicated objective function without losing their performance characteristics.

\paragraph{Definitions and Notation}

Throughout, let $G = (V, E)$ be a graph with vertex set $V$ and edge set $E$. Our graphs are
undirected, and contain no loops. Associated with $G$ is a set of labels, and we are given a mapping
from edges to labels. We are also given a budget, which is a strictly positive integer.

The \emph{neighbourhood} of a vertex is the set of vertices adjacent to it, and its \emph{degree} is
the cardinality of its neighbourhood. A \emph{colouring} of a set of vertices is an assignment of
colours to vertices, such that adjacent vertices are given different colours.  A \emph{clique} is a
set of pairwise-adjacent vertices. The \emph{cost} of a clique is the cardinality of the union of
the labels associated with all of its edges. A clique is \emph{feasible} if it has cost not greater
than the budget. We say that a feasible clique $C'$ is \emph{better than} a feasible clique $C$ if
either it has larger cardinality, or if it has the same cardinality but lower cost. The
\emph{maximum labelled clique problem} is to find a feasible clique which is either better than or
equal to any other feasible clique in a given graph---that is, of all the maximum feasible cliques,
we seek the cheapest.

The hardness of the maximum clique problem immediately implies that the maximum labelled clique
problem is also \NP-hard. Carrabs et al.\ showed that the problem remains hard even for complete
graphs, where the maximum clique problem is trivial.

\section{A Branch and Bound Algorithm}

In \cref{algorithm:main} we present the first dedicated algorithm for the maximum labelled clique
problem.  This is a branch and bound algorithm, using a greedy colouring for the bound. We start by
discussing how the algorithm finds cliques, and then explain how labels and budgets are checked.

\begin{algorithm}\DontPrintSemicolon
    \nl $\FuncSty{maximumLabelledClique}$ :: (Graph $G$, Int $\budget$) $\rightarrow$ Vertex Set \;
    \nl \Begin{
        \nl permute $G$ so that vertices are in non-increasing degree order \label{line:permute} \;
        \nl $\KwSty{global}$ ($\Cbest$, $\Lbest$) $\gets$ ($\emptyset$, $\emptyset$) \label{line:cbest} \;
        \nl $\FuncSty{expand}$($\KwSty{true}$, $\emptyset$, every vertex of $G$, $\emptyset$) \label{line:initial} \label{line:pass1} \;
        \nl $\FuncSty{expand}$($\KwSty{false}$, $\emptyset$, every vertex of $G$, $\emptyset$) \label{line:pass2} \;
        \nl $\KwSty{return}$ $\Cbest$ (unpermuted) \;
    }

    \vspace{0.5em}

    \nl $\FuncSty{expand}$ :: (Boolean $\first$, Vertex Set $C$, Vertex Set $P$, Label Set $L$) \;
    \nl \Begin{
        \nl ($\order$, $\bounds$) $\gets$ $\FuncSty{colourOrder}$($P$) \label{line:colour} \;
        \nl \For{$i$ $\gets$ $|P|$ $\KwSty{downto}$ 1 \label{line:loopstart}}{
            \nl \If{\textnormal{$|C|$ + $\bounds[i]$ $<$ $|\Cbest|$ $\KwSty{or}$
                ($\first$ $\KwSty{and}$ $|C|$ + $\bounds[i]$ $=$ $|\Cbest|$)}\label{line:bound}}{
                \nl $\KwSty{return}$}
            \nl $v$ $\gets$ $\order[i]$ \label{line:v} \;
            \nl add $v$ to $C$ \label{line:vinstart} \label{line:vtoc} \;
            \nl $L'$ $\gets$ $L$ $\cup$ the labels of edges between $v$ and any vertex in $C$ \label{line:lprime} \;
            \nl \If{\textnormal{$|L'|$ $\le$ ($\budget$ $\KwSty{if}$ $\first$, $\KwSty{otherwise}$ $|\Lbest| - 1$)}\label{line:lcheck}}{
                \nl \lIf{\textnormal{($C$, $L'$) is better than ($\Cbest$, $\Lbest$)}}{($\Cbest$, $\Lbest$) $\gets$ ($C$, $L'$) \label{line:unseat}}
                \nl $P'$ $\gets$ the vertices in $P$ that are adjacent to $v$ \label{line:pprime} \;
                \nl \lIf{$P'$ $\ne$ $\emptyset$}{$\FuncSty{expand}$($\first$, $C$, $P'$, $L'$) \label{line:recurse}}
            } \label{line:vinend}
            \nl remove $v$ from $C$ and from $P$ \label{line:vnotin} \label{line:vfromc} \label{line:loopend} \;
        }
    }

    \vspace{0.5em}

    \nl $\FuncSty{colourOrder}$ :: (Vertex Set $P$) $\rightarrow$ (Vertex Array, Int Array) \;
    \nl \Begin{
        \nl ($\order$, $\bounds$) $\gets$ ($[]$, $[]$) \;
        \nl $\uncoloured$ $\gets$ $P$ \;
        \nl $\colour$ $\gets$ $1$ \;
        \nl \While{$\uncoloured$ $\ne$ $\emptyset$\label{line:cloopoutstart}}{
            \nl $\colourable$ $\gets$ $\uncoloured$ \;
            \nl \While{$\colourable$ $\ne$ $\emptyset$\label{line:cloopstart}}{
                \nl $v$ $\gets$ the first vertex of $\colourable$ \label{line:cv} \;
                \nl append $v$ to $\order$, and $\colour$ to $\bounds$ \label{line:cgive} \;
                \nl remove $v$ from $\uncoloured$ and from $\colourable$ \;
                \nl remove from $\colourable$ all vertices adjacent to $v$ \label{line:removeadjacent}\label{line:cloopend} \;
            }
            \nl add $1$ to $\colour$ \label{line:cnewcolour}\label{line:cloopoutend}
        }
        \nl $\KwSty{return}$ ($\order$, $\bounds$) \;
    }

    \caption{An algorithm for the maximum labelled clique problem.}
    \label{algorithm:main}
\end{algorithm}

\paragraph{Branching:}

Let $v$ be some vertex in our graph. Any clique either contains only $v$ and possibly some vertices
adjacent to $v$, or does not contain $v$. Thus we may build up potential solutions by recursively
selecting a vertex, and branching on whether or not to include it. We store our growing clique in a
variable $C$, and vertices which may potentially be added to $C$ are stored in a variable $P$.
Initially $C$ is empty, and $P$ contains every vertex (line~\ref{line:initial}).

The $\FuncSty{expand}$ function is our main recursive procedure. Inside a loop
(lines~\ref{line:loopstart} to~\ref{line:loopend}), we select a vertex $v$ from $P$
(line~\ref{line:v}). First we consider including $v$ in $C$ (lines~\ref{line:vinstart}
to~\ref{line:vinend}). We produce a new $P'$ from $P$ by rejecting any vertices which are not
adjacent to $v$ (line~\ref{line:pprime})---this is sufficient to ensure that $P'$ contains only
vertices adjacent to \emph{every} vertex in $C$. If $P'$ is not empty, we may potentially grow $C$
further, and so we recurse (line~\ref{line:recurse}).  Having considered $v$ being in the clique, we
then reject $v$ (line~\ref{line:vnotin}) and repeat.

\paragraph{Bounding:}

If we can colour a graph using $k$ colours, we know that the graph cannot contain a clique of size
greater than $k$ (each vertex in a clique must be given a different colour). This gives us a bound
on how much further $C$ could grow, using only the vertices remaining in $P$. To make use of this
bound, we keep track of the largest feasible solution we have found so far (called the
\emph{incumbent}), which we store in $\Cbest$. Initially $\Cbest$ is empty (line~\ref{line:cbest}).
Whenever we find a new feasible solution, we compare it with $\Cbest$, and if it is larger, we
unseat the incumbent (line~\ref{line:unseat}).

For each recursive call, we produce a constructive colouring of the vertices in $P$
(line~\ref{line:colour}), using the $\FuncSty{colourOrder}$ function. This process produces an array
$\order$ which contains a permutation of the vertices in $P$, and an array of bounds, $\bounds$, in
such a way that the subgraph induced by the first $i$ vertices of $\order$ may be coloured using
$\bounds[i]$ colours. The $\bounds$ array is non-decreasing ($\bounds[i + 1] \ge \bounds[i]$), so if
we iterate over $\order$ from right to left, we can avoid having to produce a new colouring for each
choice of $v$. We make use of the bound on line~\ref{line:bound}: if the size of the growing clique
plus the number of colours used to colour the vertices remaining in $P$ is not enough to unseat the
incumbent, we abandon search and backtrack.

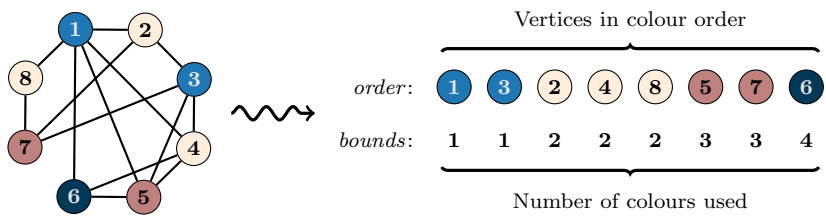
\begin{figure}[b]
    \centering
    \begin{tikzpicture}
        \input{figure-colour}
    \end{tikzpicture}
    \caption{The graph on the left has been coloured greedily: vertices 1 and 3 were given the first
    colour, then vertices 2, 4 then 8 were given the second colour, then vertices 5 and 7 were given
    the third colour, then vertex 6 was given the fourth colour. On the right, we show the $\order$
    array, which contains the vertices in the order which they were coloured. Below, the $\bounds$
    array, containing the number of colours used so far.}
    \label{figure:colour}
\end{figure}

The $\FuncSty{colourOrder}$ function performs a simple greedy colouring. We select a vertex
(line~\ref{line:cv}) and give it the current colour (line~\ref{line:cgive}). This process is
repeated until no more vertices may be given the current colour without causing a conflict
(lines~\ref{line:cloopstart} to~\ref{line:cloopend}). We then proceed with a new colour
(line~\ref{line:cnewcolour}) until every vertex has been coloured (lines~\ref{line:cloopoutstart}
to~\ref{line:cloopoutend}). Vertices are placed into the $\order$ array in the order in which they
were coloured, and the $i$th entry of the $\bounds$ array contains the number of colours used at the
time the $i$th vertex in $\order$ was coloured. This process is illustrated in \cref{figure:colour}.

\paragraph{Initial vertex ordering:}

The order in which vertices are coloured can have a substantial effect upon the colouring produced.
Here we will select vertices in a static non-increasing degree order. This is done by permuting the
graph at the top of search (line~\ref{line:permute}), so vertices are simply coloured in numerical
order. This assists with the bitset encoding, which we discuss below.

\paragraph{Labels and the budget:}

So far, what we have described is a variation of a series of maximum clique algorithms by Tomita et
al.\ \cite{Tomita:2003,Tomita:2007,Tomita:2010} (and we refer the reader to these papers to justify
the vertex ordering and selection rules chosen). Now we discuss how to handle labels and budgets. We
are optimising subject to two criteria, so we will take a two-pass approach to finding an optimal
solution.

On the first pass ($\first = \KwSty{true}$, from line~\ref{line:pass1}), we concentrate on finding
the largest feasible clique, but do not worry about finding the cheapest such clique. To do so, we
store the labels currently used in $C$ in the variable $L$. When we add a vertex $v$ to $C$, we
create from $L$ a new label set $L'$ and add to it any additional labels used
(line~\ref{line:lprime}). Now we check whether we have exceeded the budget (line~\ref{line:lcheck}),
and only proceed with this value of $C$ if we have not. As well as storing $\Cbest$, we also keep
track of the labels it uses in $\Lbest$.

On the second pass ($\first = \KwSty{false}$, from line~\ref{line:pass2}), we already have the size
of a maximum feasible clique in $|\Cbest|$, and we seek to either reduce the cost $|\Lbest|$, or
prove that we cannot do so. Thus we repeat the search, starting with our existing values of $\Cbest$
and $\Lbest$, but instead of using the budget to filter labels on line~\ref{line:lcheck}, we use
$|\Lbest| - 1$ (which can become smaller as cheaper solutions are found). We must also change the
bound condition slightly: rather than looking only for solutions strictly larger than $\Cbest$, we
are now looking for solutions with size equal to $\Cbest$ (line~\ref{line:bound}). Finally, when
potentially unseating the incumbent (line~\ref{line:unseat}), we must check to see if either $C$ is
larger than $\Cbest$, or it is the same size but cheaper.

This two-pass approach is used to avoid spending a long time trying to find a cheaper clique of size
$|\Cbest|$, only for this effort to be wasted when a larger clique is found. The additional
filtering power from having found a clique containing only one additional vertex is often extremely
beneficial. On the other hand, label-based filtering using $|\Lbest| - 1$ rather than the budget is
not possible until we are sure that $\Cbest$ cannot grow further, since it could be that larger
feasible maximum cliques have a higher cost.

\paragraph{Bit parallelism:}

For the maximum clique problem, San Segundo et al.\ \cite{SanSegundo:2011,SanSegundo:2011b} observed
that using a bitset encoding for SIMD-like parallelism could speed up an implementation by a factor
of between two to twenty, without changing the steps taken. We do the same here: $P$ and $L$ should
be bitsets, and the graph should be represented using an adjacency bitset for each vertex (this
representation may be created when $G$ is permuted, on line~\ref{line:permute}). Most importantly,
the $\uncoloured$ and $\colourable$ variables in $\FuncSty{colourOrder}$ are also bitsets, and the
filtering on line~\ref{line:removeadjacent} is simply a bitwise and-with-complement operation.

Note that $C$ should not be stored as a bitset, to speed up line~\ref{line:lprime}. Instead, it
should be an array. Adding a vertex to $C$ on line~\ref{line:vtoc} may be done by appending to the
array, and when removing a vertex from $C$ on line~\ref{line:vfromc} we simply remove the last
element---this works because $C$ is used like a stack.

\paragraph{Thread parallelism:}

Thread parallelism for the maximum clique problem has been shown to be extremely beneficial
\cite{Depolli:2013,McCreesh:2013}; we may use an approach previously described by the authors
\cite{McCreesh:2013,McCreesh:2014.shape} here too. We view the recursive calls to $\FuncSty{expand}$
as forming a tree, ignore left-to-right dependencies, and explore subtrees in parallel. For work
splitting, we initially create subproblems by dividing the tree immediately below the root node (so
each subproblem represents a case where $|C| = 1$ due to a different choice of vertex). Subproblems
are placed onto a queue, and processed by threads in order. To improve balance, when the queue is
empty and a thread becomes idle, work is then stolen from the remaining threads by resplitting the
final subproblems at distance 2 from the root.

There is a single shared incumbent, which must be updated carefully. This may be stored using an
atomic, to avoid locking. Care must be taken with updates to ensure that $\Cbest$ and $\Lbest$ are
compared and updated simultaneously---this may be done by using a large unsigned integer, and
allocating the higher order bits to $|\Cbest|$ and the lower order bits to the bitwise complement of
$|\Lbest|$.

Note that we are not dividing a fixed amount of work between multiple threads, and so we should not
necessarily expect a linear speedup. It is possible that we could get no speedup at all, due to
threads exploring a portion of the search space which would be eliminated by the bound during a
sequential run, or a speedup greater than the number of threads, due to a strong incumbent being
found more quickly \cite{Lai:1984}. A further complication is that in the first pass, we could find
an equally sized but more costly incumbent than we would find sequentially. Thus we cannot even
guarantee that this will not cause a slowdown in certain cases \cite{Trienekens:1990}.

\section{Experimental Results}

\afterpage{\clearpage\begin{landscape}
\begin{table}
    \caption{Experimental results. For each graph, we use three different label set sizes and three
        different budgets, with randomly allocated labels, and show averages over 100 runs. In each
        case, we show the average size and cost of the result, the sequential runtime in seconds, the
        parallel runtime in seconds (2 cores, 4 threads) and then the ``Enhanced'' times reported by
        Carrabs et al.\ \cite{Carrabs:2014}.}\label{table:results}

    \centering\setlength{\tabcolsep}{4.5pt}
    \begin{tabular}{lS[table-format=2] c@{\hskip 8pt}
        S[table-format=1.2]S[table-format=2.2]S[table-format=1.2]S[table-format=1.2]S[table-format=4.2] c@{\hskip 8pt}
        S[table-format=2.2]S[table-format=2.2]S[table-format=2.2]S[table-format=1.2]S[table-format=4.2] c@{\hskip 8pt}
        S[table-format=2.2]S[table-format=2.2]S[table-format=2.2]S[table-format=1.2]S[table-format=4.2]}
        \toprule
        & & &
        \multicolumn{5}{c}{25\% budget} &
            &
        \multicolumn{5}{c}{50\% budget} &
            &
        \multicolumn{5}{c}{75\% budget} \\

        \cmidrule(lr){4-8}
        \cmidrule(lr){10-14}
        \cmidrule(lr){16-20}

        \textbf{Instance} &
        $|L|$ &

        &

        {size} &
        {cost} &
        {$t_{\mathit{seq}}$} &
        {$t_{\mathit{par}}$} &
        {\cite{Carrabs:2014}} &

        &

        {size} &
        {cost} &
        {$t_{\mathit{seq}}$} &
        {$t_{\mathit{par}}$} &
        {\cite{Carrabs:2014}} &

        &

        {size} &
        {cost} &
        {$t_{\mathit{seq}}$} &
        {$t_{\mathit{par}}$} &
        {\cite{Carrabs:2014}} \\ \midrule

        \input{gen-table-problems}

        \bottomrule
    \end{tabular}
\end{table}
\end{landscape}}

We now evaluate an implementation of our sequential and parallel algorithms experimentally. Our
implementation was coded in C++, and for parallelism, C++11 native threads were used. The bitset
encoding was used in both cases. Experimental results are produced on a desktop machine with an
Intel i5-3570 CPU and 12GBytes of RAM. This is a dual core machine, with hyper-threading, so for
parallel results we use four threads (but should not expect an ideal-case speedup of 4). Sequential
results are from a dedicated sequential implementation, not from a parallel implementation run with
a single thread.  Timing results include preprocessing time and thread startup costs, but not the
time taken to read in the graph file and generate random labels.

\paragraph{Standard Benchmark Problems}

In \cref{table:results} we present results from the same set of benchmark instances as Carrabs et
al.\ \cite{Carrabs:2014}. These are some of the smaller graphs from the DIMACS implementation
challenge\footnote{\url{http://dimacs.rutgers.edu/Challenges/}}, with randomly allocated labels.
Carrabs et al.\ used three samples for each measurement, and presented the average; we use one
hundred. Note that our CPU is newer than that of Carrabs et al., and we have not attempted to scale
their results for a ``fair'' comparison.

The most significant result is that none of our parallel runtime averages are above seven seconds,
and none of our sequential runtime averages are above twenty four seconds (our worst sequential
runtime from any instance is 32.3 seconds, and our worst parallel runtime is 8.4 seconds). This is
in stark contrast to Carrabs et al., who aborted some of their runs on these instances after three
hours. Most strikingly, the keller4 instances, which all took Carrabs et al.\ at least an hour, took
under 0.1 seconds for our parallel algorithm. We are using a different model CPU, so results are not
directly comparable, but we strongly doubt that hardware differences could contribute to more than
one order of magnitude improvement in the runtimes.

We also see that parallelism is in general useful, and is never a penalty, even with very low
runtimes. We see a speedup of between 3 and 4 on the non-trivial instances. This is despite the
initial sequential portion of the algorithm, the cost of launching the threads, the general
complications involved in parallel branch and bound, and the hardware providing only two ``real''
cores.

\paragraph{Large Sparse Graphs}

In \cref{table:erdos} we present results using the Erd\H{o}s collaboration graphs from the Pajek
dataset by Vladimir Batagelj and Andrej
Mrvar\footnote{\url{http://vlado.fmf.uni-lj.si/pub/networks/data/}}. These are large, sparse graphs,
with up to 7,000 vertices (representing authors) and 12,000 edges (representing collaborations). We
have chosen these datasets because of the potential ``social network analysis'' application
suggested by Carrabs et al., where edge labels represent a particular kind of common interest, and
we are looking for a clique using only a small number of interests.

For each instance we use 3, 4 and 5 labels, with a budget of 2, 3 and 4.  The ``3 labels, budget 4''
cases are omitted, but we include the ``3 labels, budget 3'' and ``4 labels, budget 4''
cases---although the clique sizes are the same (and are equal to the size of a maximum unlabelled
clique), we see in a few instances the costs do differ where the budget is 4. Again, we use randomly
allocated labels and a sample size of 100.

Despite their size, none of these graphs are at all challenging for our algorithm, with average
sequential runtimes all being under 0.2 seconds. However, no benefit at all is gained from
parallelism---the runtimes are dominated by the cost of preprocessing and encoding the graph, not
the search.

\begin{table}[tb]
    \caption{Experimental results on Erd\H{o}s collaboration graphs. For each instance, we use three
        different label set sizes and three different budgets, with randomly allocated labels, and
        show averages over 100 runs. In each case, we show the average size and cost of the result,
        and the sequential runtime in seconds.}\label{table:erdos}

    \centering\setlength{\tabcolsep}{4.5pt}
    \begin{tabular}{lS[table-format=1] c@{\hskip 8pt}
        S[table-format=1.2]S[table-format=1.2]S[table-format=1.2] c@{\hskip 8pt}
        S[table-format=1.2]S[table-format=1.2]S[table-format=1.2] c@{\hskip 8pt}
        S[table-format=1.2]S[table-format=1.2]S[table-format=1.2]}
        \toprule
        & & &
        \multicolumn{3}{c}{budget = 2} &
        &
        \multicolumn{3}{c}{budget = 3} &
        &
        \multicolumn{3}{c}{budget = 4} \\

        \cmidrule(lr){4-6}
        \cmidrule(lr){8-10}
        \cmidrule(lr){12-14}

        \textbf{Instance} &
        $|L|$ &
        &
        {size} &
        {cost} &
        $t_{\mathit{seq}}$ &
        &
        {size} &
        {cost} &
        $t_{\mathit{seq}}$ &
        &
        {size} &
        {cost} &
        $t_{\mathit{seq}}$ \\ \midrule

        \input{gen-table-problems2}

        \bottomrule
    \end{tabular}
\end{table}

\section{Possible Improvements and Variations}

We will briefly describe three possible improvements to the algorithm. These have all been
implemented and appear to be viable, but for simplicity we do not go into detail on these points. We
did not use these improvements for the results in the previous section. We also suggest a variation
of the problem.

\paragraph{Resuming where we left off:} Rather than doing two full passes, it is possible to start
the second pass at the point where the last unseating of the incumbent occurred in the first pass.
In the sequential case, this is conceptually simple but messy to implement: viewing the recursive
calls to $\FuncSty{expand}$ as a tree, we could store the location whenever the incumbent is
unseated.  For the second pass, we could then skip portions of the search space ``to the left'' of
this point.  In parallel, this is much trickier: it is no longer the case that when a new incumbent
is found, we have necessarily explored every subtree to the left of its position.

\paragraph{Different initial vertex orders:} We order vertices by non-increasing degree order at the
top of search. Other vertex orderings have been proposed for the maximum clique problem, including a
dynamic degree and exdegree ordering \cite{Tomita:2010}, and minimum-width based orderings
\cite{Prosser:2012,SanSegundo:2014}. These orderings give small improvements for the harder problem
instances when labels are present. However, for the Erd\H{o}s graphs, dynamic degree and exdegree
orderings were a severe penalty---they are more expensive to compute (adding almost a whole second
to the runtime), and the search space is too small for this one-time cost to be ignored.

\paragraph{Reordering colour classes:} For the maximum clique problem, small but consistent benefits
can be had by permuting the colour class list produced by $\FuncSty{colourOrder}$ to place colour
classes containing only a single vertex at the end, so that they are selected first
\cite{McCreesh:2014.reorder}. A similar benefit is obtained by doing this here.

\paragraph{A multi-label variation of the problem:} In the formulation by Carrabs et al., each edge
has exactly one label. What if instead edges may have multiple labels? If taking an edge requires
paying for all of its labels, this is just a trivial modification to our algorithm. But if taking an
edge requires selecting and paying for only one of its labels, it is not obvious what the best way
to handle this would be. One possibility would be to branch on edges as well as on vertices (but
only where none of the available edges matches a label which has already been selected).

This modification to the problem could be useful for real-world problems: for Carrabs et al.'s
example where labels represent different relationship types in a social network graph, it is
plausible that two people could both be members of the same club and be colleagues.  Similarly, for
the Erd\H{o}s datasets, we could use labels either for different journals and conferences, or for
different topic areas (combinatorics, graph theory, etc.). When looking for a clique of people using
only a small number of different relationship types, it would make sense to allow only one of the
relationships to count towards the cost. However, we suspect that this change could make the problem
substantially more challenging.

\section{Conclusion}

We saw that our dedicated algorithm was faster than a mathematical programming solution. This is not
surprising. However, the extent of the performance difference was unexpected: we were able to solve
multiple problems in under a tenth of one second that previously took over an hour, and we never
took more than ten seconds to solve any of Carrabs et al.'s instances. We were also able to work
with large sparse graphs without difficulty.

Of course, a more complicated mathematical programming model could close the performance gap. One
possible route, which has been successful for the maximum clique problem in a SAT setting
\cite{Li:2011}, would be to treat colour classes as variables rather than vertices. But this would
require a pre-processing step, and would lose the ``ease of use'' benefits of a mathematical
programming approach. It is also not obvious how the label constraints would map to this kind of
model, since equivalently coloured vertices are no longer equal.

On the other hand, adapting a dedicated maximum clique algorithm for this problem did not require
major changes. It is true that these algorithms are non-trivial to implement, but there are at least
three implementations with publicly available source code (one in Java \cite{Prosser:2012} and two
with multi-threading support in C++ \cite{Depolli:2013,McCreesh:2013}). Also of note was that bit-
and thread-parallelism, which are key contributors to the raw performance of maximum clique
algorithms, were similarly successful in this setting.

A further surprise is that threading is beneficial even with the low runtimes of some problem
instances. We had assumed that our parallel runtimes would be noticeably worse for extremely easy
instances, but this turned out not to be the case. Although there was no benefit for the Erd\H{o}s
collaboration graphs, which were computationally trivial, for the DIMACS graphs there were clear
benefits from parallelism even with sequential runtimes as low as a tenth of a second. For the
non-trivial instances, we consistently obtained speedups of between 3 and 4. Even on inexpensive
desktop machines, it is worth making use of multiple cores.

\bibliographystyle{spmpsci}
\bibliography{maximum_labelled_clique}

\end{document}

%% file: figure-clique.tex

\tikzstyle{l1} = [color=safeverydarkblue];
\tikzstyle{l2} = [color=safemediumorange];
\tikzstyle{l3} = [ultra thick, color=safedarkblue];
\tikzstyle{l4} = [ultra thick, color=safelightorange];

\newcount \c
\foreach \n in {1, ..., 7}{
    \c=\n \advance\c by -1 \multiply\c by 360 \divide\c by 7 \advance\c by 90
    \node[draw, circle, fill=white, inner sep=2pt, font=\small] (N\n) at (\the\c:1.2) {\textbf{\n}};
}

\draw [l4] (N3) -- (N4);
\draw [l4] (N1) -- (N3);
\draw [l2] (N1) -- (N5);
\draw [l2] (N2) -- (N3);
\draw [l2] (N2) -- (N4);
\draw [l2] (N3) -- (N5);
\draw [l2] (N4) -- (N5);
\draw [l2] (N4) -- (N6);
\draw [l2] (N5) -- (N7);
\draw [l1] (N1) -- (N2);
\draw [l1] (N2) -- (N5);
\draw [l3] (N4) -- (N7);
\draw [l3] (N5) -- (N6);
\draw [l3] (N6) -- (N7);
\draw [l3] (N1) -- (N4);

\draw [l1] (1.9,  0.6) -- (2.3,  0.6); \node [font=\small, anchor = west] at (2.35,  0.6) { $l_1$ };
\draw [l2] (1.9,  0.2) -- (2.3,  0.2); \node [font=\small, anchor = west] at (2.35,  0.2) { $l_2$ };
\draw [l3] (1.9, -0.2) -- (2.3, -0.2); \node [font=\small, anchor = west] at (2.35, -0.2) { $l_3$ };
\draw [l4] (1.9, -0.6) -- (2.3, -0.6); \node [font=\small, anchor = west] at (2.35, -0.6) { $l_4$ };

%% file: figure-colour.tex

\tikzstyle{l} = [thick];

\newcount \c
\foreach \n in {1, ..., 8}{
    \c=\n \advance\c by -1 \multiply\c by -360 \divide\c by 8 \advance\c by 90 \advance\c by 22.5
    \ifthenelse{\n = 1 \OR \n = 3}{
        \node[draw, circle, fill=safedarkblue, text=black!10!white, inner sep=2pt, font=\small] (N\n) at (\the\c:1.2) {\textbf{\n}};
    }{}
    \ifthenelse{\n = 2 \OR \n = 4 \OR \n = 8} {
        \node[draw, circle, fill=safelightorange, inner sep=2pt, font=\small] (N\n) at (\the\c:1.2) {\textbf{\n}};
    }{}
    \ifthenelse{\n = 5 \OR \n = 7}{
        \node[draw, circle, fill=safemediumorange, inner sep=2pt, font=\small] (N\n) at (\the\c:1.2) {\textbf{\n}};
    }{}
    \ifthenelse{\n = 6}{
        \node[draw, circle, fill=safeverydarkblue, text=black!10!white, inner sep=2pt, font=\small] (N\n) at (\the\c:1.2) {\textbf{\n}};
    }{}
}

\draw [l] (N1) -- (N2);
\draw [l] (N1) -- (N4);
\draw [l] (N1) -- (N5);
\draw [l] (N1) -- (N6);
\draw [l] (N1) -- (N8);
\draw [l] (N2) -- (N3);
\draw [l] (N2) -- (N7);
\draw [l] (N3) -- (N4);
\draw [l] (N3) -- (N5);
\draw [l] (N3) -- (N7);
\draw [l] (N4) -- (N5);
\draw [l] (N4) -- (N6);
\draw [l] (N5) -- (N6);
\draw [l] (N7) -- (N8);

\draw [->, very thick, decorate, decoration={snake, post length=0.5mm}] (1.6, 0) -> (2.7, 0);

\coordinate (Ms) at (4.3, 0.35);
\node[right = 0.0 of Ms, draw, circle, fill=safedarkblue, text=black!10!white, inner sep=2pt, font=\small]     (M1) {\textbf{1}};
\node[right = 0.2 of M1, draw, circle, fill=safedarkblue, text=black!10!white, inner sep=2pt, font=\small]     (M2) {\textbf{3}};
\node[right = 0.2 of M2, draw, circle, fill=safelightorange, inner sep=2pt, font=\small]  (M3) {\textbf{2}};
\node[right = 0.2 of M3, draw, circle, fill=safelightorange, inner sep=2pt, font=\small]  (M4) {\textbf{4}};
\node[right = 0.2 of M4, draw, circle, fill=safelightorange, inner sep=2pt, font=\small]  (M5) {\textbf{8}};
\node[right = 0.2 of M5, draw, circle, fill=safemediumorange, inner sep=2pt, font=\small] (M6) {\textbf{5}};
\node[right = 0.2 of M6, draw, circle, fill=safemediumorange, inner sep=2pt, font=\small] (M7) {\textbf{7}};
\node[right = 0.2 of M7, draw, circle, fill=safeverydarkblue, text=black!10!white, inner sep=2pt, font=\small] (M8) {\textbf{6}};

\node[left = 0.2 of M1, font=\small] {$\order$:};
\draw[decorate, decoration={brace}, very thick] ($(M1.north west)+(0.0,0.3)$) -- ($(M8.north east)+(0.0,0.3)$);
\node[anchor=south, font=\small] at ($(M1)!0.5!(M8)$)[yshift=0.7cm] { Vertices in colour order };

\coordinate (Bs) at (4.3, -0.35);
\node[right = 0.0 of Bs, draw, circle, fill=white, color=white, text=black, inner sep=2pt, font=\small] (B1) {\textbf{1}};
\node[right = 0.2 of B1, draw, circle, fill=white, color=white, text=black, inner sep=2pt, font=\small] (B2) {\textbf{1}};
\node[right = 0.2 of B2, draw, circle, fill=white, color=white, text=black, inner sep=2pt, font=\small] (B3) {\textbf{2}};
\node[right = 0.2 of B3, draw, circle, fill=white, color=white, text=black, inner sep=2pt, font=\small] (B4) {\textbf{2}};
\node[right = 0.2 of B4, draw, circle, fill=white, color=white, text=black, inner sep=2pt, font=\small] (B5) {\textbf{2}};
\node[right = 0.2 of B5, draw, circle, fill=white, color=white, text=black, inner sep=2pt, font=\small] (B6) {\textbf{3}};
\node[right = 0.2 of B6, draw, circle, fill=white, color=white, text=black, inner sep=2pt, font=\small] (B7) {\textbf{3}};
\node[right = 0.2 of B7, draw, circle, fill=white, color=white, text=black, inner sep=2pt, font=\small] (B8) {\textbf{4}};

\node[left = 0.2 of B1, font=\small] {$\bounds$:};
\draw[decorate, decoration={brace}, very thick] ($(B8.south east)+(0.0,-0.2)$) -- ($(B1.south west)+(0.0,-0.2)$);
\node[anchor=north, font=\small] at ($(B1)!0.5!(B8)$)[yshift=-0.6cm] { Number of colours used };

%% file: gen-table-problems.tex
\multicolumn{1}{l}{\textbf{johnson8-2-4}} & 
4 & & 
3.13  & 1.00  & 0.00 & 0.00 & \color{gray}0.03&&
4.00  & 1.87  & 0.00 & 0.00 & \color{gray}0.02&&
4.00  & 1.87  & 0.00 & 0.00 & \color{gray}0.01\\
\textcolor{gray}{\hspace{1em}28 vertices,} & 
8 & & 
3.51  & 1.50  & 0.00 & 0.00 & \color{gray}0.03&&
4.00  & 2.48  & 0.00 & 0.00 & \color{gray}0.01&&
4.00  & 2.48  & 0.00 & 0.00 & \color{gray}0.02\\
\textcolor{gray}{\hspace{1em}210 edges} & 
12 & & 
4.00  & 2.85  & 0.00 & 0.00 & \color{gray}0.01&&
4.00  & 2.85  & 0.00 & 0.00 & \color{gray}0.02&&
4.00  & 2.85  & 0.00 & 0.00 & \color{gray}0.02\\[0.1cm]
\multicolumn{1}{l}{\textbf{MANN\_a9}} & 
11 & & 
5.64  & 2.76  & 0.01 & 0.00 & \color{gray}3.18&&
8.89  & 5.93  & 0.02 & 0.01 & \color{gray}13.98&&
13.34  & 8.99  & 0.01 & 0.00 & \color{gray}6.47\\
\textcolor{gray}{\hspace{1em}45 vertices,} & 
21 & & 
6.61  & 5.60  & 0.02 & 0.01 & \color{gray}12.43&&
9.74  & 10.68  & 0.08 & 0.02 & \color{gray}46.52&&
13.32  & 15.73  & 0.03 & 0.01 & \color{gray}15.55\\
\textcolor{gray}{\hspace{1em}918 edges} & 
32 & & 
7.00  & 7.79  & 0.04 & 0.01 & \color{gray}17.61&&
10.26  & 14.99  & 0.19 & 0.05 & \color{gray}108.13&&
14.12  & 23.28  & 0.03 & 0.01 & \color{gray}13.18\\[0.1cm]
\multicolumn{1}{l}{\textbf{hamming6-4}} & 
6 & & 
3.99  & 1.97  & 0.00 & 0.00 & \color{gray}0.32&&
4.00  & 1.99  & 0.00 & 0.00 & \color{gray}0.33&&
4.00  & 1.99  & 0.00 & 0.00 & \color{gray}0.35\\
\textcolor{gray}{\hspace{1em}64 vertices,} & 
11 & & 
4.00  & 2.64  & 0.00 & 0.00 & \color{gray}0.43&&
4.00  & 2.64  & 0.00 & 0.00 & \color{gray}0.42&&
4.00  & 2.64  & 0.00 & 0.00 & \color{gray}0.37\\
\textcolor{gray}{\hspace{1em}704 edges} & 
17 & & 
4.00  & 2.98  & 0.00 & 0.00 & \color{gray}0.41&&
4.00  & 2.98  & 0.00 & 0.00 & \color{gray}0.43&&
4.00  & 2.98  & 0.00 & 0.00 & \color{gray}0.44\\[0.1cm]
\multicolumn{1}{l}{\textbf{hamming6-2}} & 
15 & & 
6.07  & 3.82  & 0.04 & 0.01 & \color{gray}41.32&&
9.87  & 7.86  & 0.76 & 0.21 & \color{gray}382.73&&
15.42  & 12.00  & 2.44 & 0.66 & \color{gray}711.20\\
\textcolor{gray}{\hspace{1em}64 vertices,} & 
29 & & 
7.17  & 7.12  & 0.41 & 0.11 & \color{gray}221.77&&
11.01  & 14.73  & 6.25 & 1.70 & \color{gray}2368.78&&
15.86  & 21.84  & 11.73 & 3.14 & \color{gray}2079.50\\
\textcolor{gray}{\hspace{1em}1824 edges} & 
44 & & 
8.00  & 10.81  & 1.35 & 0.37 & \color{gray}429.21&&
12.00  & 21.49  & 20.19 & 5.47 & \color{gray}4955.96&&
17.13  & 32.65  & 23.15 & 6.28 & \color{gray}2170.84\\[0.1cm]
\multicolumn{1}{l}{\textbf{johnson8-4-4}} & 
14 & & 
5.99  & 3.93  & 0.01 & 0.00 & \color{gray}49.82&&
7.98  & 6.94  & 0.02 & 0.01 & \color{gray}85.73&&
11.09  & 10.89  & 0.01 & 0.00 & \color{gray}16.36\\
\textcolor{gray}{\hspace{1em}70 vertices,} & 
27 & & 
6.30  & 5.95  & 0.03 & 0.01 & \color{gray}102.60&&
9.07  & 12.81  & 0.02 & 0.01 & \color{gray}110.53&&
12.18  & 20.29  & 0.00 & 0.00 & \color{gray}11.80\\
\textcolor{gray}{\hspace{1em}1855 edges} & 
40 & & 
7.01  & 8.97  & 0.05 & 0.01 & \color{gray}177.13&&
10.00  & 19.01  & 0.02 & 0.01 & \color{gray}60.79&&
12.97  & 28.93  & 0.00 & 0.00 & \color{gray}2.39\\[0.1cm]
\multicolumn{1}{l}{\textbf{johnson16-2-4}} & 
23 & & 
6.50  & 5.28  & 0.15 & 0.04 & \color{gray}4248.29&&
8.00  & 8.91  & 0.14 & 0.04 & \color{gray}1943.33&&
8.00  & 8.91  & 0.14 & 0.04 & \color{gray}2066.26\\
\textcolor{gray}{\hspace{1em}120 vertices,} & 
46 & & 
7.75  & 11.17  & 0.29 & 0.08 & \color{gray}5660.29&&
8.00  & 12.21  & 0.22 & 0.06 & \color{gray}3044.75&&
8.00  & 12.21  & 0.22 & 0.06 & \color{gray}3290.52\\
\textcolor{gray}{\hspace{1em}5460 edges} & 
69 & & 
8.00  & 14.23  & 0.28 & 0.08 & \color{gray}3699.71&&
8.00  & 14.23  & 0.28 & 0.08 & \color{gray}4227.76&&
8.00  & 14.23  & 0.28 & 0.08 & \color{gray}3909.15\\[0.1cm]
\multicolumn{1}{l}{\textbf{keller4}} & 
28 & & 
6.98  & 6.89  & 0.28 & 0.07 & {\textcolor{gray}{$>3h$}}&&
9.04  & 12.68  & 0.10 & 0.03 & {\textcolor{gray}{$>3h$}}&&
11.00  & 18.75  & 0.02 & 0.01 & \color{gray}3304.30\\
\textcolor{gray}{\hspace{1em}171 vertices,} & 
55 & & 
8.00  & 12.85  & 0.32 & 0.09 & {\textcolor{gray}{$>3h$}}&&
11.00  & 26.98  & 0.02 & 0.01 & \color{gray}4081.97&&
11.00  & 26.98  & 0.02 & 0.01 & \color{gray}4173.34\\
\textcolor{gray}{\hspace{1em}9435 edges} & 
83 & & 
9.00  & 19.82  & 0.14 & 0.04 & {\textcolor{gray}{$>3h$}}&&
11.00  & 31.88  & 0.02 & 0.01 & \color{gray}4827.99&&
11.00  & 31.88  & 0.02 & 0.01 & \color{gray}5028.16\\[0.1cm]

%% file: gen-table-problems2.tex
\multicolumn{1}{l}{\textbf{Erdos971}} & 
3 & & 
5.20 & 1.98 & 0.00 &&
7.00 & 3.00 & 0.00 &&
& &
\\
\textcolor{gray}{\hspace{1em}472 vertices,} & 
4 & & 
4.61 & 1.86 & 0.00 &&
5.71 & 2.78 & 0.00 &&
7.00 & 3.98 & 0.00 \\
\textcolor{gray}{\hspace{1em}1314 edges} & 
5 & & 
4.24 & 1.94 & 0.00 &&
5.10 & 2.81 & 0.00 &&
6.05 & 3.91 & 0.00 \\[0.1cm]
\multicolumn{1}{l}{\textbf{Erdos972}} & 
3 & & 
5.30 & 1.98 & 0.12 &&
7.00 & 3.00 & 0.12 &&
& &
\\
\textcolor{gray}{\hspace{1em}5488 vertices,} & 
4 & & 
4.84 & 1.90 & 0.12 &&
5.84 & 2.88 & 0.12 &&
7.00 & 3.97 & 0.12 \\
\textcolor{gray}{\hspace{1em}8972 edges} & 
5 & & 
4.35 & 1.80 & 0.12 &&
5.18 & 2.76 & 0.12 &&
6.04 & 3.82 & 0.12 \\[0.1cm]
\multicolumn{1}{l}{\textbf{Erdos981}} & 
3 & & 
5.18 & 1.99 & 0.00 &&
7.00 & 3.00 & 0.00 &&
& &
\\
\textcolor{gray}{\hspace{1em}485 vertices,} & 
4 & & 
4.68 & 1.91 & 0.00 &&
5.78 & 2.88 & 0.00 &&
7.00 & 3.99 & 0.00 \\
\textcolor{gray}{\hspace{1em}1381 edges} & 
5 & & 
4.18 & 1.82 & 0.00 &&
5.21 & 2.89 & 0.00 &&
6.10 & 3.82 & 0.00 \\[0.1cm]
\multicolumn{1}{l}{\textbf{Erdos982}} & 
3 & & 
5.29 & 1.99 & 0.14 &&
7.00 & 3.00 & 0.14 &&
& &
\\
\textcolor{gray}{\hspace{1em}5822 vertices,} & 
4 & & 
4.80 & 1.85 & 0.14 &&
5.95 & 2.95 & 0.14 &&
7.00 & 3.96 & 0.14 \\
\textcolor{gray}{\hspace{1em}9505 edges} & 
5 & & 
4.26 & 1.69 & 0.14 &&
5.19 & 2.82 & 0.14 &&
6.19 & 3.88 & 0.14 \\[0.1cm]
\multicolumn{1}{l}{\textbf{Erdos991}} & 
3 & & 
5.18 & 1.97 & 0.00 &&
7.00 & 3.00 & 0.00 &&
& &
\\
\textcolor{gray}{\hspace{1em}492 vertices,} & 
4 & & 
4.64 & 1.89 & 0.00 &&
5.84 & 2.91 & 0.00 &&
7.00 & 3.98 & 0.00 \\
\textcolor{gray}{\hspace{1em}1417 edges} & 
5 & & 
4.23 & 1.93 & 0.00 &&
5.19 & 2.82 & 0.00 &&
6.16 & 3.86 & 0.00 \\[0.1cm]
\multicolumn{1}{l}{\textbf{Erdos992}} & 
3 & & 
5.36 & 2.00 & 0.15 &&
8.00 & 3.00 & 0.15 &&
& &
\\
\textcolor{gray}{\hspace{1em}6100 vertices,} & 
4 & & 
4.92 & 1.92 & 0.15 &&
6.06 & 2.98 & 0.15 &&
8.00 & 3.99 & 0.15 \\
\textcolor{gray}{\hspace{1em}9939 edges} & 
5 & & 
4.27 & 1.84 & 0.15 &&
5.32 & 2.86 & 0.15 &&
6.38 & 3.88 & 0.15 \\[0.1cm]
\multicolumn{1}{l}{\textbf{Erdos02}} & 
3 & & 
5.86 & 2.00 & 0.19 &&
8.00 & 3.00 & 0.19 &&
& &
\\
\textcolor{gray}{\hspace{1em}6927 vertices,} & 
4 & & 
5.04 & 1.99 & 0.19 &&
6.43 & 2.98 & 0.19 &&
8.00 & 3.99 & 0.19 \\
\textcolor{gray}{\hspace{1em}11850 edges} & 
5 & & 
4.66 & 1.82 & 0.19 &&
5.69 & 2.83 & 0.19 &&
6.82 & 3.91 & 0.19 \\[0.1cm]